\documentclass[aps,twocolumn,preprintnumbers,%
superscriptaddress,floatfix]{revtex4}
\usepackage[dvips]{graphics}
\usepackage{amsmath,amsfonts,bm}
\usepackage{epsfig}
\begin{document}
\newcommand{\be}{\begin{eqnarray}}
\newcommand{\ee}{\end{eqnarray}}
\def\lsim{\mathrel{\rlap{\lower3pt\hbox{\hskip1pt$\sim$}}
     \raise1pt\hbox{$<$}}} 
\def\gsim{\mathrel{\rlap{\lower3pt\hbox{\hskip1pt$\sim$}}
     \raise1pt\hbox{$>$}}} 
\def\N{${\cal N}\,\,$}
\def\prl{Phys. Rev. Lett.}
\def\np{Nucl. Phys.}
\def\pr{Phys. Rev.}
\def\pl{Phys. Lett.}
\def\la{\langle}\def\ra{\rangle}
\def\del{\partial}
\def\calL{\cal L}\def\calK{\cal K}
\def\hatn{\hat{n}}\def\Amu{{\cal A}_\mu}\def\A{{\cal A}}
\newcommand\<{\langle}
\renewcommand\>{\rangle}
\renewcommand\d{\partial}
\newcommand\LambdaQCD{\Lambda_{\textrm{QCD}}}
\newcommand\Tr{\mathrm{Tr}\,}
\newcommand\+{\dagger}
\newcommand\g{g_5}
\def\bi{\bibitem}

\newcommand{\msun}{\mbox{$M_\odot$}}
\newcommand{\rsun}{\mbox{$R_\odot$}}

\title{Dileptons Get Nearly ``Blind" to Mass-Scaling Effects In Hot and/or Dense Matter}
\author{Mannque Rho}
 \affiliation{ Institut de Physique Th\'eorique,  CEA Saclay, 91191 Gif-sur-Yvette C\'edex, France\\
 and Department of Physics, Hanyang University, 133-791 Seoul, Korea}

\begin{abstract}
Based on BHHRS~\cite{BHHRS} and further sharpened by discussions with Gerry Brown that I had in October 2008, we~\cite{footnote-we-i} arrive at the present assessment of the dilepton saga, namely, that dileptons become ``blind" to changes in the vacuum structure of chiral symmetry (such as, e.g., BR scaling~\cite{BR91}) at high temperature and/or at high density and hence are {\em not} an appropriate probe for a signal for partial or complete chiral restoration, contrary to what has been widely believed. There, however, are a variety of indirect indications that the scaling notion is qualitatively, if not quantitatively, valid and should work in various low-energy nuclear phenomena, and it is fair to conclude that while there is no direct evidence for the scaling notion, there is {\em none against} it either, in disagreement with the claim made in \cite{cern,green}. I will touch briefly on certain observables that could give a clear-cut litmus signal for the vacuum structure of chiral symmetry modified by temperature and/or density.
\end{abstract}

\date{\today}

\newcommand\sect[1]{\emph{#1}---}

\maketitle
\sect{$\bullet$ The upshot}
 There seems to be a general consensus in the heavy-ion community~\cite{cern} (1) that the dilepton spectra of the recent refined measurements as NA60 and others can be described more or less fully by {\em conventional hadronic many-body effects} alone~\cite{footnote1}, and (2) that these results imply  there can be little, if any,  ``mass-shift" of hadrons caused by the change of vacuum structure of chiral symmetry (CVCS for short) in hot and/or dense matter.  We agree with the first statement but disagree strongly with the second. Our assertion is that the dileptons measured in the experiments performed so far are {\em nearly blind} to those vector mesons which carry information on chiral symmetry properties of hot and/or dense medium. I would like to clarify what our (Gerry's and my) points are as they stand now.

As in BHHRS~\cite{footnote2}, we suggest that dileptons are {\em strongly}, though perhaps not completely, suppressed in the temperature regime defined as ``hadronic freedom" (HF for short) between the chiral transition temperature $T_c$ and the flash temperature $T_{flash}$ in which the CVCS effect should be discernable and hence whatever carries information on chiral symmetry is swamped by the mundane effects that I shall simply refer to as ``garbages"~\cite{footnote2p}. So within the largely uncontrollable uncertainty in theory as well as in dealing with experimental data (such as evolution code, ``cocktails" etc), the dilepton measurements so far performed could not properly single out the CVCS that one is looking for. It would be grossly unjustified if not totally wrong to conclude that the CVCS was absent in the measured process.

\sect{$\bullet$ CVCS in nuclei}
There are a variety of evidences in nuclear physics that the CVCS is present and operative. But they are all indirect. In some sense, the situation is very similar to meson-exchange currents in nuclei or at the most fundamental level, to the Lamb shift in QED. In order to ``see" the effect in question, all possible mundane nuclear effects that are not {\em directly}~\cite{footnote2pp} connected to the CVCS (i.e., ``garbages") have to be accurately identified and subtracted out. In the case of meson-exchange currents, this required that all possible nucleonic effects involving many-body dynamics be accurately calculated within one well-defined and consistent theoretical scheme (e.g., chiral perturbation theory) before meson-exchange effects could be singled out of the same calculation~\cite{MR}. It took many decades since Yukawa's work on meson theory for this effect to be confirmed.

Let me just cite a couple of recent cases to illustrate my point. Other examples are given in review articles by Gerry and me~\cite{BR-PR1,BR-PR2,BR-PR3} and also in my recent book~\cite{MR-book}.
\begin{enumerate}
\item The C-14 dating: This is one of the most remarkable cases where a simple or even what I would consider a ``naive" accounting of BR scaling is revealed rather spectacularly. It comes in the tensor force contributed by the $\rho$ exchange between two nucleons in nuclear matter. That nuclear tensor forces get a substantial suppression in nuclear matter was pointed out in 1990~\cite{BR-tensor}. The mechanism is primarily that the $\rho$ tensor force which comes with the sign opposite to that of the pion tensor force gets enhanced effectively by the factor $(g^*/m_\rho^*)^2$ (here $g$ is the $\rho NN$ coupling related to hidden gauge coupling in hidden local symmetry (HLS)~\cite{HY}) where the asterisk denotes in-medium quantity. Now it has been established that $g^*\simeq g$ up to $n\sim n_0$ where $n_0$ is the normal nuclear matter density, whereas $m_\rho^*/m_\rho\simeq f_\pi^*/f_\pi\approx 1-\kappa (n/n_0)$ where $\kappa$ can be somewhere between 0.1 and 0.2. This effect is used by Holt et al~\cite{holt1} to explain the striking effect in the C-14 dating  which exploits that the process occurs below but very near the nuclear matter density $n_0$. Interestingly it has been suggested that the C-14 dating could also be explained by three-body forces~\cite{holt2}. I will suggest below that these mechanisms are related.
\item The nuclear symmetry energy $S$ crucial for neutron star structure: It has been suggested that the symmetry energy that fits the $\pi^-/\pi^+$ ratio of FOPI/GSI~\cite{FOPI} could be understood in terms of the same BR-scaled $\rho$ tensor force that figures in the C-14 dating~\cite{BAL}. To be certain, it would require further confirmation, both experimental and theoretical, that the present FOPI/GSI does indeed constrain $S$ in the way claimed in \cite{BAL1}. The verdict will presumably come from FAIR/GSI or even from the RIB accelerators in construction~\cite{BAL-private}. But let me assume that it's OK, focusing on qualitative aspects involved. Remarkably, as suggested in \cite{BAL}, it can also be explained by the three-body forces of the sort that figure in the C-14 dating. The FOPI data appear to constrain $S$ only up to about $\sim$ 1.3$n_0$, and beyond that density, nothing is precisely known. Now in Ref.\cite{BAL1}, the authors assume that $(g^*/m_\rho^*)^2\simeq (g/m_\rho^*)^2$ for {\em all} $n$ beyond $n_0$. We have suggested -- based on hidden local symmetry with vector manifestation (HLS/VM for short)~\cite{footnote1} -- that $(g^*/m_\rho^*)\approx {\rm const.}$ for $n > n_{flash}$  where $n_{flash}$ density must be a bit above the maximum density measured by the FOPI data.  In terms of three-body forces, this feature would correspond to the {\em suppression} of three-body force effects at some density above $\sim 1.3n_0$. This means that the analysis made in \cite{BAL1} could very well be invalidated by this HLS/VM effect. My guess is that the non-Newtonian gravity effect considered in \cite{BAL1} will not be indispensable for reconciling the symmetry energy $S$ with neutron-star data once HLS/VM scaling is taken into account.
\end{enumerate}

\sect{$\bullet$ BR scaling vs. three-body forces}
At this point, let me make a conjecture. It's only a conjecture for the moment but it needs to be made more rigorous.

The conjecture is that doing chiral perturbation calculation including up to three-body forces with a chiral Lagrangian whose parameters are determined in the vacuum is equivalent -- in some given channels -- to doing chiral perturbation calculation {\em without} three-body forces with the same chiral Lagrangian whose parameters are determined at the Landau Fermi liquid fixed point, i.e., BR scaling fit to the pion decay constant at $n\simeq n_0$.  Song~\cite{song} has shown how this connection via what's called ``rearrangement terms" is required for thermodynamic consistency of many-body dynamics. Now what about for densities $n > n_0$? I would suggest that the HLS/VM effect discussed above in the two-body case, that is, $m_\rho^*$ scaling along with $g^*$,  captures the mechanism of the three-body forces getting {\em suppressed} at high density. One could see this in terms of  $\omega$-exchange three-body forces accounting for the contact three-body term with the coefficient $C_E$ in \cite{holt1}. There is a similar 5D dual interpretation of this suppression in holographic QCD~\cite{hashimoto}. It is shown there that at short distances, one can use flat geometry in 5D in which case the ADHM multi-instanton solution is directly applicable for many-nucleon dynamics. One can therefore analytically calculate N-body forces and show that they are suppressed as $V_N \sim {\cal O}(N_c/\lambda^{N-1})$ where $\lambda$ is the 't Hooft constant. Since in hadronic world, $N_c/\lambda\sim {\cal O} (10^{-1})$, N-body forces will be strongly suppressed {\em at short distances}. I think this can be translated into the VM in 4D.

\sect{$\bullet$ CVCS in dilepton production}
Let me consider the following HLS scenario. Consider an HLS theory in which baryons are introduced via topology, i.e., skyrmions, and scalars are introduced via the trace anomaly as recently discussed in \cite{HKL-MR}, namely in terms of the soft dilaton $\chi_s$. Let me endow the parameters of this Lagrangian, which I will just call ${\cal L}_{HLS}^\prime$, with a suitable scaling (like BR scaling for the mass and coupling as alluded above). Now compute the $\rho$ spectral function with this Lagrangian taking into account thermal and dense loops including all relevant degrees of freedom. In computing this spectral function, I would have to do the thermal evolution from, say, $T_c$, to $T_{flash}$ at which $\sim 90\%$ on-shell masses are recovered, and the strong interactions recover nearly full on-shell strengths. Let's follow the temperature evolution going down from $T_c$. I will not consider what happens above $T_c$.
\begin{enumerate}
\item From  $T_c$ to $T_{flash}$ (``hadronic freedom" (HF) regime), there will be BR-scaled sharp low-mass dileptons as dictated by HLS/VM, but we argue that they are {\em strongly} suppressed by HLS/VM as in BHHRS.~\cite{footnote4}
\item At $T_{flash}$, all hadrons entering in the process, as discussed in connection with the STAR $\rho^0/\pi^-$ ratio~\cite{rho-pi}, go on-shell with their strong coupling strengths recovered. All the variety of different nuclear processes, such as couplings to sobars, collisional broadening etc. taken into account in various different conventional many-body hadronic approaches (generically referred to as $nonVM$~\cite{footnote1}) are essentially getting the same thing since the parameters of all the $nonVM$ models, though different in specific contents, are {\em fit} to on-shell data. We don't have our own numerical codes constructed {\em in consistency} with ${\cal L}_{HLS}^\prime$ etc, so cannot compute numerically, and compare with, what the experimentalists call the $\rho$ spectral function. However if our theory ${\cal L}_{HLS}^\prime$ is to give on-shell quantities correctly -- which it certainly should~\cite{HY}, then it must describe correctly the spectral function {\em appropriate for the flash point}. I don't see how HLS/VM can go wrong on-shell including many-body processes. We in fact looked at the NA60 spectrum -- modulo absolute normalization -- predicted by ${\cal L}_{HLS}^\prime$ and we were satisfied that the $\rho$ spectrum, including the shoulder below the $\rho$ peak, can be understood entirely (at least qualitatively) by mundane processes taking place on-shell at and below $T_{flash}$ assuming that the dileptons coming from the $\rho$ mesons produced in the HF regime are highly suppressed. Furthermore, whatever dileptons that may have been produced from the $\rho$ mesons in the HF regime (hence BR scaled) will appear below the $\rho$ peak, but, highly suppressed, they will be swamped by the on-shell production (involving width broadened by mundane nuclear effects, p-wave pion-pion scattering etc.) and will be buried in the error bars there. They will indeed be ``needles" in the haystack as we have been saying all along.
\end{enumerate}
What could make a $nonVM$ calculation numerically different from what we will get from our scenario is, however, that in the $nonVM$ scenario, dileptons are produced vector-dominated with on-shell conditions {\em throughout} the evolution, {\em including the narrow HF regime.} From $T_{flash}$ downwards, there will be no essential differences between the two theories since it's the on-shell action that will figure for which our model should give, when fully calculated, the same result as any good phenomenological model which is fit to data in the matter-free vacuum would do.

So where would the difference show up? The answer is that it will be in the absolute normalization.

{\it If it can be shown without any doubt that the missing strength between the temperature range between $T_c$ and $T_{flash}$ implied by HLS/VM  is inconsistent with the experimental absolute normalization, then that will be a clear falsification of the HLS/VM scenario}:  In HLS/VM, the dilepton yield from the $\rho$ within the HF regime will be mostly missing due to the suppression in that regime.

However in the presence of so many free parameters and fudge factors in the phenomenological fitting, it is highly doubtful that whatever differences there may be in the absolute normalization due to what happens in a narrow temperature regime where CVCS figures -- which is expected to be insignificant -- could be discerned. In this sense, there cannot be any noticeable differences between what the $nonVM$-type calculation gets and what we predict with a consistent ${\cal L}_{HLS}^\prime$ theory. Thus as it stands, {\em the dileptons measured so far can say nothing about CVCS!!}

\sect{$\bullet$ ``Seeing" BR scaling}
What can one do with the measured dilepton yields to ``see" the effect of CVCS?

Here is one way. Subtract away all ``garbages" from the dilepton yields: That is, subtract not only all the usual cocktails but also all dileptons emitted from ``on-shell" vector mesons. For the latter, both HLS/VM with ${\cal L}^\prime_{HLS}$ and realistic phenomenological models -- there may be more than one in the literature -- in the $nonVM$ class will do equally well. What result will then be (a) the dileptons directly coupled to pions that populate the HF regime, this coupling being present due to the strong in-medium violation of vector dominance in HLS/VM, plus (b) those dileptons with (suppressed) coupling to the sharp BR-scaled $\rho$ mesons expected in HLS/VM, namely, those ``needles" in the haystack with the haystack removed. The dilepton coupling to pions in the HF regime will be point-like, so will be structureless. This feature will be totally absent in the $nonVM$ scenario where HLS/VM or BR scaling is missing. Perhaps this feat is much too daunting a task to perform.

\sect{$\bullet$ ``Smoking gun" signal} The argument that dileptons are ``blind" to BR-scaling vector mesons applies equally to dense medium. In fact, as stressed elsewhere~\cite{BR-PR2,BR-PR3}, the HF regime sets in a lot more precociously in dense matter than in hot medium. This is already manifest in that vector dominance in the sense of Sakurai's is badly violated in the nucleon form factor already in matter free space. It fact the photon couples to nucleon roughly half of the time via vector mesons and the other half directly. In holographic QCD, this is understood by the fact that the nucleon form factor is vector-dominated by the {\em infinite tower} of vector mesons -- and not by the lowest members -- and it is thus a bad approximation for nucleon form factors to couple the photon 100\% to the lowest vector mesons $\rho$ and $\omega$~\cite{HRYY}. This will be more so in dense matter: Density will make dileptons {\em even more blind} to BR-scaling vector mesons than temperature does. Thus the CEBAF's negative result is of no great surprise.

But then is there any process that can show a clear evidence for HLS/VM scaling?

Here is one smoking-gun evidence for or against HLS/VM. Measure the pion velocity $v_\pi$ at or near $T_c$. If there is no massless or no near massless $\rho$ meson, then one should find $v_\pi\approx 0$ as predicted by Son and Stephanov~\cite{son-stephanov}. $v_\pi\approx 0$ will kill HLS/VM and hence BR scaling. However if there is massless $\rho$ near $T_c$, then one should find $v_\pi\approx 1$~\cite{HKRS}. $v_\pi\approx 1$ will then be an unequivocal support for HLS/VM as well as for BR scaling. This is a night-and-day difference. There may be other signals and they should be looked for.

\sect{Acknowledgments} This work was supported by the WCU project of Korean Ministry of Education, Science and Technology (R33-2008-000-10087-0).


\begin{thebibliography}{99}
\bibitem{footnote-we-i} In what follows, ``we" will stand for Gerry Bown and myself. Otherwise ``I" will be used throughout.
\bibitem{footnote2} Here I focus on temperature effect but a similar argument holds for density. I will also deal with the chiral limit although symmetry breaking effects are not always ignorable.
\bibitem{footnote2p} Or broadly, part of the background to the phenomenon one is to zero in.
\bibitem{footnote4} Note, however, that in this HF regime, due to VD violation, dileptons of various invariant masses will be produced by direct point-like coupling to pions which will not enter in the proper $\rho$ spectral function. In holographic dual description in 5D, the point-like coupling corresponds to the infinite tower above the ground state vector mesons. I will come back to this point below.
\end{thebibliography}
\end{document}